# Quantum Hall Effect in Black Phosphorus Two-dimensional Electron Gas

Likai Li[1,4], Fangyuan Yang[1,4], Guo Jun Ye[2,3,4], Zuocheng Zhang[5], Zengwei Zhu[6], Wen-Kai Lou[7,8], Liang Li[6], Kenji Watanabe[9], Takashi Taniguchi[9], Kai Chang[7,8], Yayu Wang[5], Xian Hui Chen[2,3,4]* and Yuanbo Zhang[1,4]*

[1]*State Key Laboratory of Surface Physics and Department of Physics, Fudan University, Shanghai 200433, China.*

[2]*Hefei National Laboratory for Physical Science at Microscale and Department of Physics, University of Science and Technology of China, Hefei, Anhui 230026, China.*

[3]*Key Laboratory of Strongly Coupled Quantum Matter Physics, University of Science and Technology of China, Hefei, Anhui 230026, China.*

[4]*Collaborative Innovation Center of Advanced Microstructures, Nanjing 210093, China.*

[5]*Department of Physics, State Key Laboratory of Low Dimensional Quantum Physics, Tsinghua University, Beijing 100084, China.*

[6]*Wuhan National High Magnetic Field Center and School of Physics, Huazhong University of Science and Technology, Wuhan 430074, China.*

[7]*SKLSM, Institute of Semiconductors, Chinese Academy of Sciences, P.O. Box 912, Beijing 100083, China*

[8]*Synergetic Innovation Center of Quantum Information and Quantum Physics, University of Science and Technology of China, Hefei, Anhui 230026, China*

[9]*Advanced Materials Laboratory, National Institute for Materials Science, 1-1 Namiki, Tsukuba, 305-0044, Japan*

*Email: zhyb@fudan.edu.cn, chenxh@ustc.edu.cn

**Development of new, high quality functional materials has been at the forefront of condensed matter research. The recent advent of two-dimensional black phosphorus has greatly enriched the material base of two-dimensional electron systems. Significant progress has been made to achieve high mobility black phosphorus two-dimensional electron gas (2DEG) since the development of the first black phosphorus field-effect transistors (FETs)[1–4]. Here, we reach a milestone in developing high quality black phosphorus 2DEG – the observation of integer quantum Hall (QH) effect. We achieve high quality by embedding the black phosphorus 2DEG in a van der Waals heterostructure close to a graphite back gate; the graphite gate screens the impurity potential in the 2DEG, and brings the carrier Hall mobility up to 6000 $cm^2V^{-1}s^{-1}$. The exceptional mobility enabled us, for the first time, to observe QH effect, and to gain important information on the energetics of the spin-split Landau levels in black phosphorus. Our results set the stage for further study on quantum transport and device application in the ultrahigh mobility regime.**

Quantum Hall effect, the emergence of quantized Hall resistance in 2DEG sample when subjected to low temperatures and strong magnetic fields, has had a lasting impact in modern condensed matter research. The exact, universal quantization regardless of detailed sample geometry and impurity configuration has enabled the establishment of a metrological resistance standard, and served as the basis for an independent determination of the fine structure constant[5]. Even though the exact quantization of the Hall resistance relies on certain amount of impurities[6], the observation of QH effect, paradoxically, requires high-purity, low-defect specimens. Because of the stringent requirement on the material quality, the effect has only been produced in a handful of materials, with each of them exhibiting distinctive features in the QH regime as a result of their peculiar electronic properties[5,7–11].

Black phosphorus 2DEG, a recent addition to the 2DEG family, brings new possibilities. The bulk black phosphorus crystal has a direct, anisotropic band gap that is predicted to increases from ~ 0.3 eV to ~ 2 eV when the material is thinned down to monolayer[12–16]. The tunable direct bandgap couples black phosphorus strongly to light in the spectrum range that spans from far-infrared to red[17–20]. Such optoelectronic properties, combined with black phosphorus's sensitivity to external perturbations such as strain/pressure and electric field[21–25], may lead to interesting new phenomena in the QH regime.

Even though bulk crystal black phosphorus was shown to exhibit mobilities on the order of ~ $10^4$ cm$^2$V$^{-1}$s$^{-1}$ (ref. 16), impurities at the interface with hexagonal boron nitride (hBN) substrate has so far limited the low-temperature Hall mobility in few-layer black phosphorus FETs to ~ $10^3$ cm$^2$V$^{-1}$s$^{-1}$ (refs. 26–30) (samples on SiO$_2$ substrate have even lower mobility[1–4]). In this work, we achieved high Hall mobility in black phosphorus FETs that is significantly higher than previous record value. This is accomplished by constructing a van der Waals heterostructure with the few-layer black phosphorus sandwiched between two hBN flakes (Fig. 1a,b) and placed on graphite back gate. The top hBN protects the black phosphorus flakes from sample degradation in air. More importantly, the thin bottom hBN (thickness ~ 25 nm) allows the electrons in the graphite to screen the impurity potential at the black phosphorus-hBN interface, where the 2DEG resides. The high mobility enable us, for the first time, to observe the QH effect in black phosphorus 2DEG. Black phosphorus thus joins the selected few materials[5,7,8,11] to become the only 2D atomic crystal apart from graphene[9,10] having requisite material quality to show QH effect.

We constructed the van der Waals heterostructure using the dry-transfer technique described in ref. 31. We first cleaved graphite and h-BN flakes onto $SiO_2$/Si wafers, and black phosphorus flake onto poly-propylene carbon (PPC) film. The black phosphorus flake on the PPC film was then used to pick up the h-BN flake on the SiO2/Si wafer. Finally, the black phosphorus/h-BN stack was released onto the graphite flake supported on $SiO_2$/Si wafer to complete the assembly of the black phosphorus/h-BN/graphite heterostructure. The polymer-assisted dry transfer ensured that there was no contamination from polymer or solvent to the black phosphorus/h-BN or the h-BN/graphite interfaces, which are crucial to the quality of the 2DEG. In addition, the transfers were performed inside a glove box with $O_2$ and $H_2O$ content below 1 ppm to avoid degradation of black phosphorus surface. We then fabricated electrodes (Cr/Au with thicknesses of 2 nm and 60 nm, respectively) on black phosphorus using standard electron-beam lithography. A second layer of h-BN was then transferred on top (not shown in Fig. 1a) to protect it from degradation during brief exposures to air between measurements.

The heterostructure significantly improves the hole carrier mobility in our black phosphorus 2DEG. Here the carrier mobility refers to the Hall mobility $\mu_H$ which serves a more reliable metric of the sample quality (field-effect mobility inflates the mobility value as a result of the carrier-density dependence of the mobility in black phosphorus 2DEG[26]). $\mu_H = \sigma/n_H e$, where $\sigma$ is the sample sheet conductance, $n_H$ is the areal carrier density obtained from Hall measurement, and $e$ is the charge of an electron. Fig. 1c displays the Hall mobility measured as a function of temperature for varying hole concentrations. The Hall mobility reaches up to $6000\ \text{cm}^2\text{V}^{-1}\text{s}^{-1}$, and saturates at this value at temperatures $T < 30\ \text{K}$ with the areal hole density fixed at $n_H = 7.0 \times 10^{12} \text{cm}^{-2}$. Such mobility considerably exceeds previous record value in black phosphrus 2DEG on hBN substrate, where no local graphite gate was available to screen the charged impruties[26–30]. The behavior of the temperature-dependent mobility, which is typical of semiconductors[32], reveals the charge transport mechanism in our black phosphorus 2DEG. Whereas the charge transport is dominated by phonons at high temperatures ($T > 30$ K; manifested as the sharp decrease in $\mu_H$), impurity scattering is still the limiting factor of the mobility. The high low-temperature mobility, however, points to much reduced influence from the charged impurities in our 2DEG. We note that gate-induced carriers themselves provide extra screening of the impurities, as our data show suppressed mobilities at reduced carrier densities at low temperatures. The high mobility of our sample is further corroborated by probing the Shubnikov-de Haas (SdH) oscillations in the magnetoresistance $R_{xx}$. Such oscillations are clearly resolved starting from a critical magnetic field of $B_c \sim 2$ T (Supplementary Section 2). An independent estimation of the carrier mobility $\mu \sim 1/B_c$ (ref. 33), therefore, puts the mobility at $\sim 5000\ \text{cm}^2\text{V}^{-1}\text{s}^{-1}$. This estimation agrees with the Hall mobility, and is further supported by the long carrier quantum lifetime obtained from the same sample (Supplementary Section 2). Improvement of carrier mobility was also clearly seen at electron doping (Supplementary Section 3), although high contact resistance at electron doping under high magnetic fields prevented a detailed study of electron carriers' quantum transport.

The exceptional carrier mobility enables us to observe the QH effect in the extreme quantum limit. Fig. 2a shows the Hall resistance $R_{xy}$ and $R_{xx}$ as a function of magnetic field $B$ at varying hole-doping levels (gate voltage $V_g < 0$; data at electron doping are discussed in Supplementary Section 3). At high magnetic fields, $R_{xy}$ exhibits quantized plateaus at $(\nu e^2/h)^{-1}$ (where integer $\nu$ is the filling factor in usual QH language) accompanied by vanishing $R_{xx}$, which are the hallmark of the QH effect. The plateaus are well-defined at $\nu = 2, 3$ and $4$, and developing plateaus are observed up to $\nu = 8$ before the QH features transform into Shubnikov-de Haas (SdH) oscillations at lower magnetic fields. Alternatively, QH effect can be probed by varying $V_g$ while fixing the magnetic field as shown in Fig. 2b. Such measurement further reveals the lowest $\nu = 1$ QH state at the extreme quantum limit, even though the large contact resistance at low doping level makes a precise measurement difficult (see Supplementary Section 4). The fact that QH states develop at both even and odd $\nu$ means that the two-fold spin degeneracy is fully lifted.

Two energy scales determine the energy gaps of the QH states at even and odd filling factors: cyclotron energy $E_c$ ($= \hbar eB/m^*$, where $\hbar$ is the reduced Plank constant, $m^*$ the carrier effective mass) and Zeeman energy $E_z$ ($= g\mu_B B$, where $g$ is the Landé $g$-factor, $\mu_B$ the Bohr magneton). Measurement of the QH gaps therefore offers precious insight into the energetics of the Landau levels (LLs) in black phosphorus 2DEG. We determine the energy gap $\Delta E$ at $\nu = 3$ and $4$ by monitoring the thermally activated behavior of $R_{xx}$ minimum, $R_{xx}^{min} \sim \exp(-\Delta E/2k_B T)$ ($k_B$ is the Boltzmann constant), at elevated temperatures (Fig. 3). Line fits to the data yield energy gaps of $\Delta E_{\nu=3} = 32.4$ K and $\Delta E_{\nu=4} = 30.9$ K at magnetic fields 27.5 T and 20.6 T, respectively (Fig. 3, left inset). Assuming a constant LL broadening $\Gamma$, the sequence of the spin-split LLs (Fig. 3 inset) dictates that $\Delta E_{odd} = E_z - 2\Gamma$ and $\Delta E_{even} = E_c - E_z - 2\Gamma$. So our determination of $\Delta E_{\nu=3}$ and $\Delta E_{\nu=4}$ provide a measure of $E_c$ and $E_z$, pending a separate estimation of $\Gamma$. One such estimation comes from $\Gamma \sim \hbar/\tau_q$, where $\tau_q = 0.36$ ps is the carrier quantum lifetime that we obtain from the Dingle plot of the low-field SdH oscillations (Supplementary Section 2). We therefore estimate that $E_c$ and $E_z$ are approximately $6.3B$ K and $2.7B$ K, respectively. We note that the Zeeman energy is an appreciable fraction of the cyclotron energy (~0.4; the fraction at electron doping should be even higher[26]). This is in contrast with graphene or GaAs-based 2DEG, where $E_z/E_c$ is on the order of $10^{-2}$, but rather close to ZnO system with an $E_z/E_c$ ratio of 0.95 (ref. 34). In light of recent observation of peculiar even-denominator fractional QH states in ZnO (ref. 34), black phosphorus may provide a new venue for exploring exotic many-body physics that are potentially relevant for topological quantum computing.

In view of the importance of the energetics of spin-split LLs, we now accurately determine the magnitude of Zeeman energy with respect to cyclotron energy. This is achieved by tilting the magnetic field by an angle $\theta$ away from the 2DEG normal (Fig. 4a inset), and monitor the SdH oscillation amplitude at odd and even filling factors. Fig. 4a presents the SdH oscillations from filling factor 5 to 15 recorded at varying $\theta$, with the total magnetic field fixed at 20 T. The relative amplitude of the oscillations between odd and even filling factors, $\Delta R_{odd}/\Delta R_{even}$, depends sensitively on their corresponding energy gaps, and thus serves as a good indicator of the relative magnitude of $\Delta E_{odd}$ and $\Delta E_{even}$ (Fig. 4b) (ref. 35). Because $E_z$ scales with total magnetic field $B_t$, whereas $E_c$ depends only on the magnetic field perpendicular to the 2DEG, $B_p = B_t \cos(\theta)$, it's possible to reach a coincidence condition, where $\Delta E_{odd} = \Delta E_{even}$, at certain critical angle, $\theta_c$. Such critical angle is readily identified with high precision in Fig. 4b as the angle where $\Delta R_{odd}/\Delta R_{even}$ crosses 1. This allows us to obtain an $E_z/E_c$ ratio of 0.345 with an uncertainty down to $\pm$ 1.8% in black phosphorus 2DEG subjected to a perpendicular magnetic field. Finally, we extract the *g*-factor based on the accurate determination of $E_z/E_c$ ratio. Because $\mu_B = \hbar e/2m_0$, where $m_0$ is the mass of a bare electron, $E_z/E_c$ takes the form $m^*g/2m_0$. (Here we assume an isotropic *g*-factor because the spin-orbit coupling is negligible in black phosphorus[36]; the assumption is validated by our explicit *g*-factor calculation presented in Supplementary Section 7) Using the effective hole mass $m^* = 0.34 \pm 0.02 \, m_0$ measured under a comparable magnetic field[26], we obtain $g = 2.0 \pm 0.1$, in good agreement with our theoretical calculations (Supplementary Section 7). We note, however, that different values of hole mass have been reported (see Supplementary Section 5 for a complete list), which complicates the determination of $g$. This may indicate that multiple factors are at play in determining the carrier effective mass, and further experimental/theoretical efforts are needed to clarify this issue.

In conclusion, we have achieved exceptionally high carrier mobility in black phosphorus 2DEG by using a graphite back gate to screen the charged impurities in the system. The high mobility enabled us, for the first time, to observe the integer QH effect in black phosphorus. Investigation of the spin-split LL energetics reveals that Zeeman energy is a significant fraction of the cyclotron energy in black phosphorus 2DEG, so that spin-split LL cross under a moderately tilted magnetic field. The large Zeeman energy relative to cyclotron energy, coupled with black phosphorus's anisotropic electronic structure, may lead to exotic quantum states in the fractional QH regime.

**Acknowledgements**

We thank Alex Hamilton, Li Yang for helpful discussions. We also thank Scott Hannahs, Tim Murphy, Eun Sang Choi, David Graf, Jonathan Billings, Bobby Pullum, Luis Balicas for help with measurements in DC high magnetic fields, Junfeng Wang, Zhengcai Xia for help with measurements in pulsed magnetic fields, and Philip Kim, Xiaomeng Liu, Lei Wang for help with the dry-transfer technique. A portion of this work was performed at the National High Magnetic Field Laboratory, which is supported by National Science Foundation Cooperative Agreement no. DMR-1157490, the State of Florida, and the U.S. Department of Energy. Measurements in pulsed magnetic field were carried out at Wuhan National High Magnetic Field Center, China. Part of the sample fabrication was conducted at Fudan Nano-fabrication Lab. L.L., F. Y. and Y.Z. acknowledge financial support from the National Basic Research Program of China (973 Program; grant nos. 2011CB921802 and 2013CB921902), and NSF of China (grant no. 11425415). L.L. and Y.Z. also acknowledge support from Samsung Global Research Outreach (GRO) Program. G.J.Y and X.H.C. acknowledge support from the 'Strategic Priority Research Program' of the Chinese Academy of Sciences (grant no. XDB04040100) and the National Basic Research Program of China (973 Program; grant no. 2012CB922002). Z.Z. and Y.W. are supported by Ministry of Science and Technology of China (grant no. 2015CB921000). W.-K.L. and K.C. acknowledge support from NSF of China (grant no. 11434010). K.W. and T.T. acknowledge support from the Elemental Strategy Initiative conducted by the MEXT, Japan. T.T. also acknowledges support by a Grant-in-Aid for Scientific Research on Innovative Areas, "Nano Informatics" (grant nos. 262480621 and 25106006) from JSPS.

## Author contributions

L.L. fabricated black phosphorus devices, performed transport measurements, and analyzed the data. F. Y. and Z. Z. helped with the transport measurement. G.J.Y. and X.H.C. grew bulk black phosphorus crystals. Z.Z. and L.L. helped with measurements in pulsed high magnetic field. W.-K.L. and K.C. did theoretical calculations. K.W. and T. T. grew bulk h-BN. Y.Z., X.H.C and Y.W. co-supervised the project. L.L. and Y.Z. wrote the paper with input from all authors.

**Figure captions**

**Figure 1 | Device structure and mobility characterisation of black phosphorus 2DEG.
a,** Optical image of a black phosphorus/h-BN/graphite heterostructure with graphite serving as the back gate. Boundaries of h-BN and graphite area are marked by green and red broken lines, respectively. A layer of h-BN (not shown) is later deposited on top to protect black phosphorus from degradation in air. **b,** Schematic three-dimensional view of the complete heterostructure stack of the device in **a**. **c,** Hall mobility $\mu_H$ as a function of temperature measured at varying hole carrier densities. Vertical error bars represent uncertainties in determining the sheet conductance $\sigma$ from the measured sample resistance as a result of the irregular sample geometry.

**Figure 2 | QH effect in black phosphorus 2DEG a,** Hall resistance (upper panel) and magnetoresistance (lower panel) as a function of magnetic field measured at varying hole doping levels ($V_g < 0$). Data were taken at $T = 300$ mK. Horizontal lines in the upper panel mark the quantized values at $h/\nu e^2$. Magnetoresistance curves are shifted vertically by multiples of 1 kΩ for clarity, and broken lines indicate zero-resistance level at each gate voltage. The QH effect in black phosphorus 2DEG is shown by at least three well-quantized plateaus in $R_{xy}$ at $\nu = 2, 3$ and 4, with vanishing $R_{xx}$ at the corresponding magnetic fields. Developing plateaus at $\nu$ up to 8 are also resolved. **b,** Hall resistance (black) and magnetoresistance (red) as a function of gate voltage. Data were obtained at $B = 31$ T and $T = 300$ mK. Integers indicate the filling factor $\nu$ at each QH states. QH plateaus are observed at $\nu$ from 1 to 6.

**Figure 3 | Measurement of QH energy gaps at $\nu = 3$ and 4.** Magnetoresistance as a function of magnetic field recorded at increasing temperatures 0.3 K (purple), 1.5 K (blue), 4 K (green), 5.1 K (yellow), 7.2 K (orange), 9 K (red) and 11.7 K (brown). Gate voltage is fixed at $-2.5$ V during measurement. Left inset: $R_{xx}$ minimum, $R_{xx}^{min}$, plotted as a function of $1/T$ in log-scale (also known as the Arrhenius plot), for $\nu = 3$ (black) and 4 (red). Both data sets fall on straight lines, indicating well-behaved thermally activated charge transport. Linear fits of the data (solid lines) yield QH energy gaps of 32.4 K and 30.9 K at $\nu = 3$ and 4, respectively. Right inset: sequence of spin-split Landau levels for hole carriers.

**Figure 4 | Probing Landau level energetics in tilted magnetic field. a,** Magnetoresistance (after subtracting a smooth background) as a function of $V_g/\cos(\theta)$ showing the amplitude of the SdH oscillations. Data were recorded in a fixed total magnetic field of 20 T while the tilt angle $\theta$ is varied. Inset: measurement configuration. Temperature was kept at 300 mK during measurement. Vertical lines mark the minima of the SdH oscillations and their corresponding filling factors $\nu$ from 5 to 15. **b,** The ratio between the SdH oscillation amplitude at odd and even filling factors, $\Delta R_{odd}/\Delta R_{even}$, as a function $V_g/\cos(\theta)$ at varying tilt angles. Data were extracted from **a**. $\Delta R_{even}$ was taken as the average of two adjacent even-filling-factor amplitudes to compensate for a slight gate-dependence of the amplitude. Critical angle $\theta_c$ ($= 46.3° \pm 1°$) is identified as the $\theta$ where $\Delta R_{odd}/\Delta R_{even}$ passes 1, which allows for an accurate determination of $E_z/E_c = 0.345 \pm 0.006$ (see text).

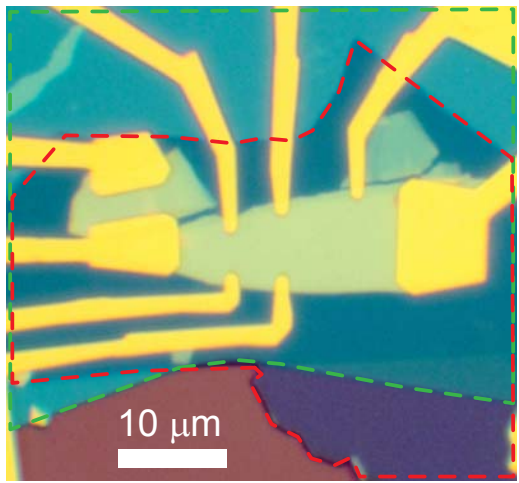
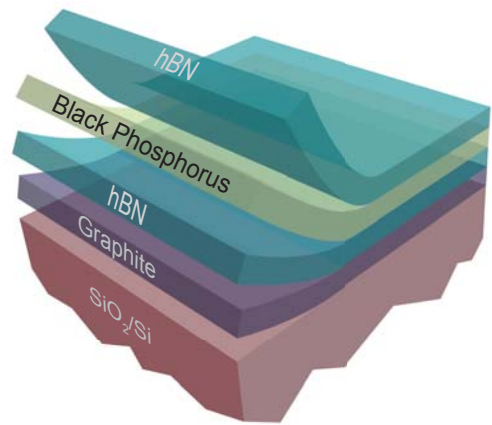
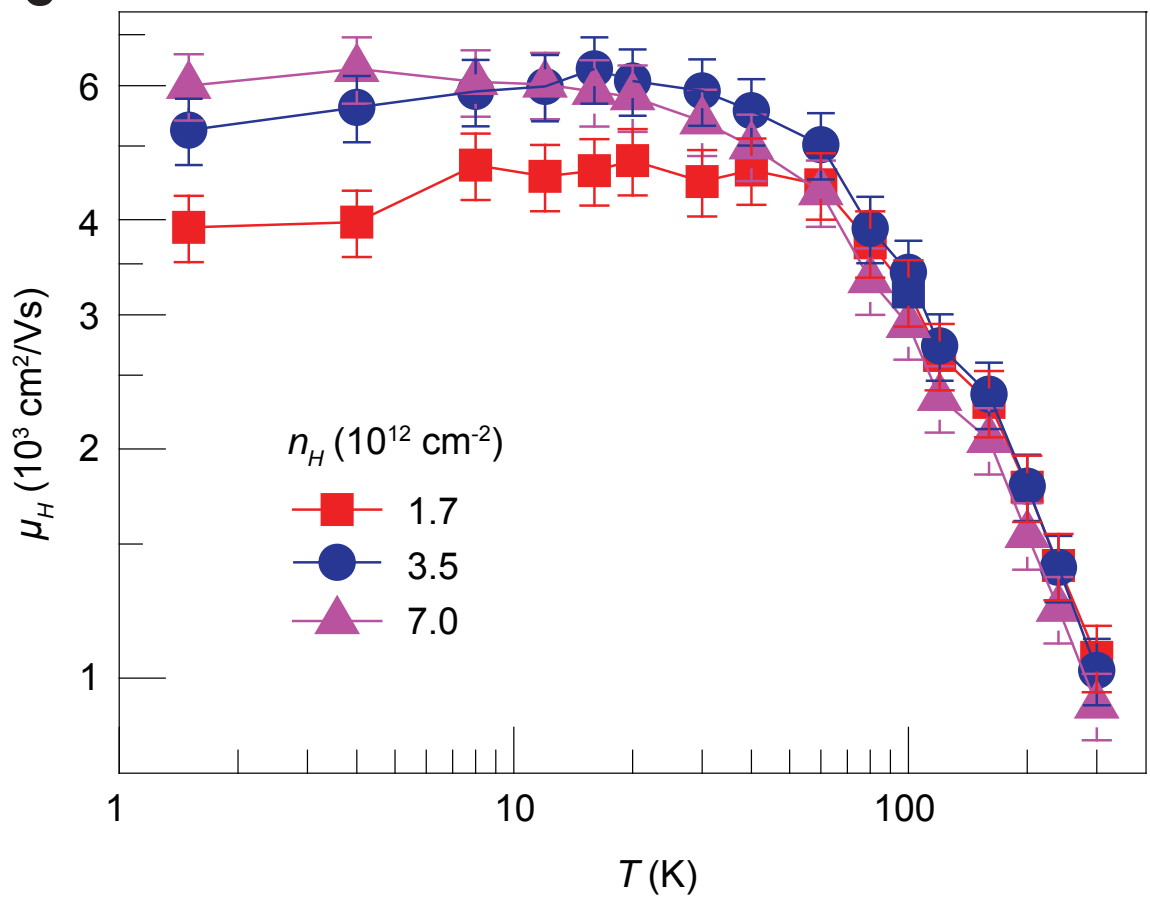

Likai Li *et al.*    Figure 1

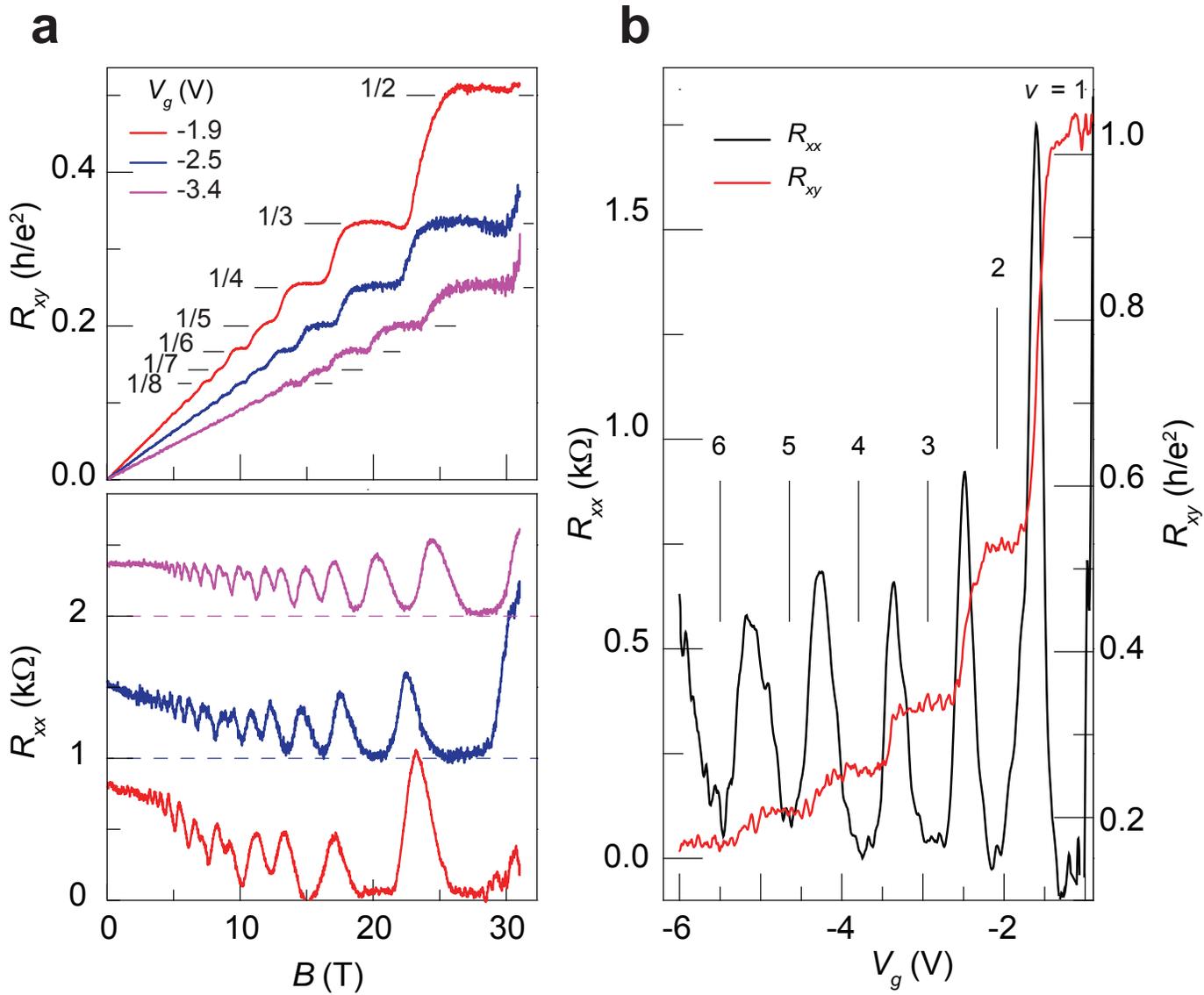

Likai Li *et al.* Figure 2

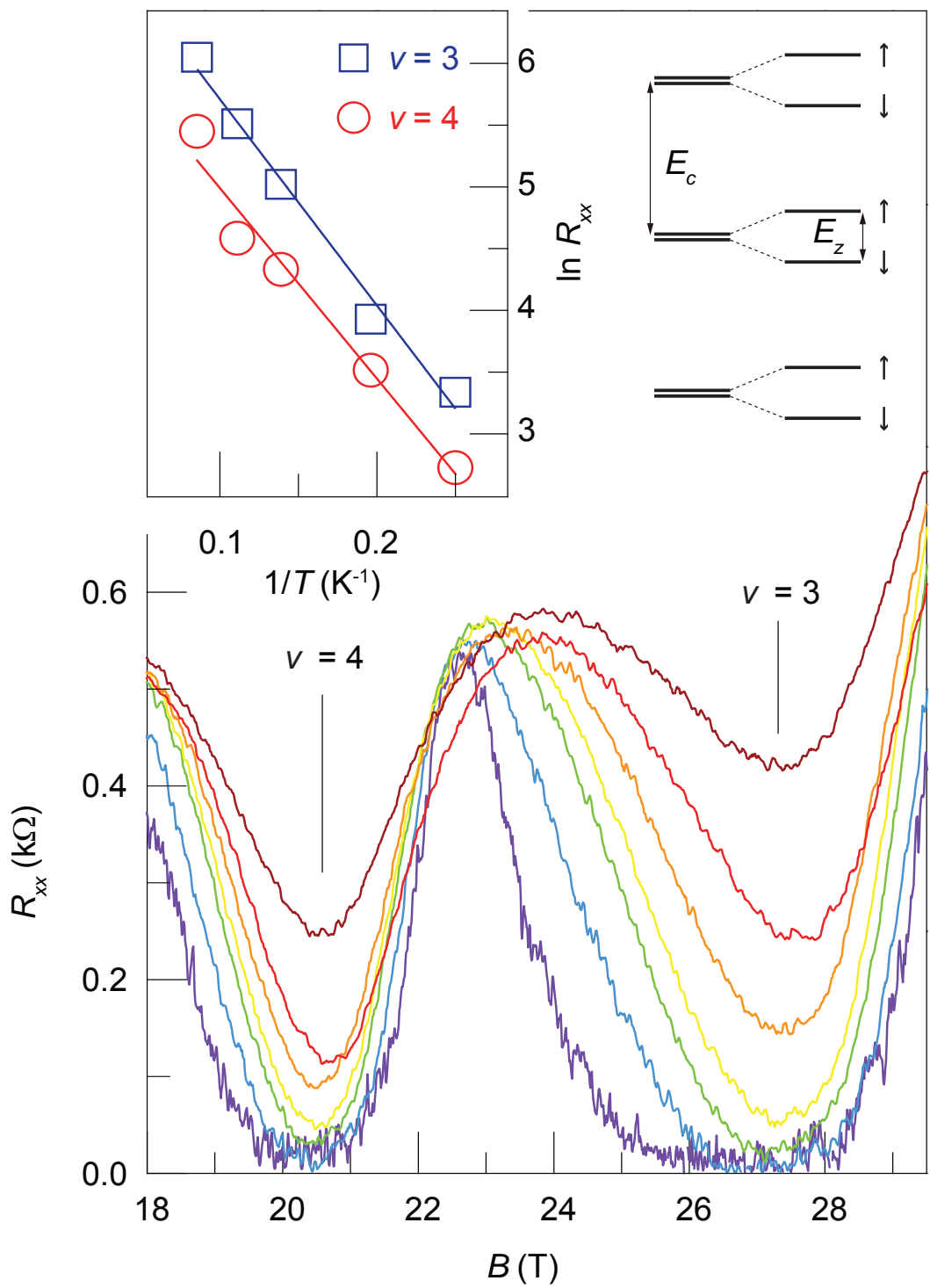

Likai Li *et al.*    Figure 3

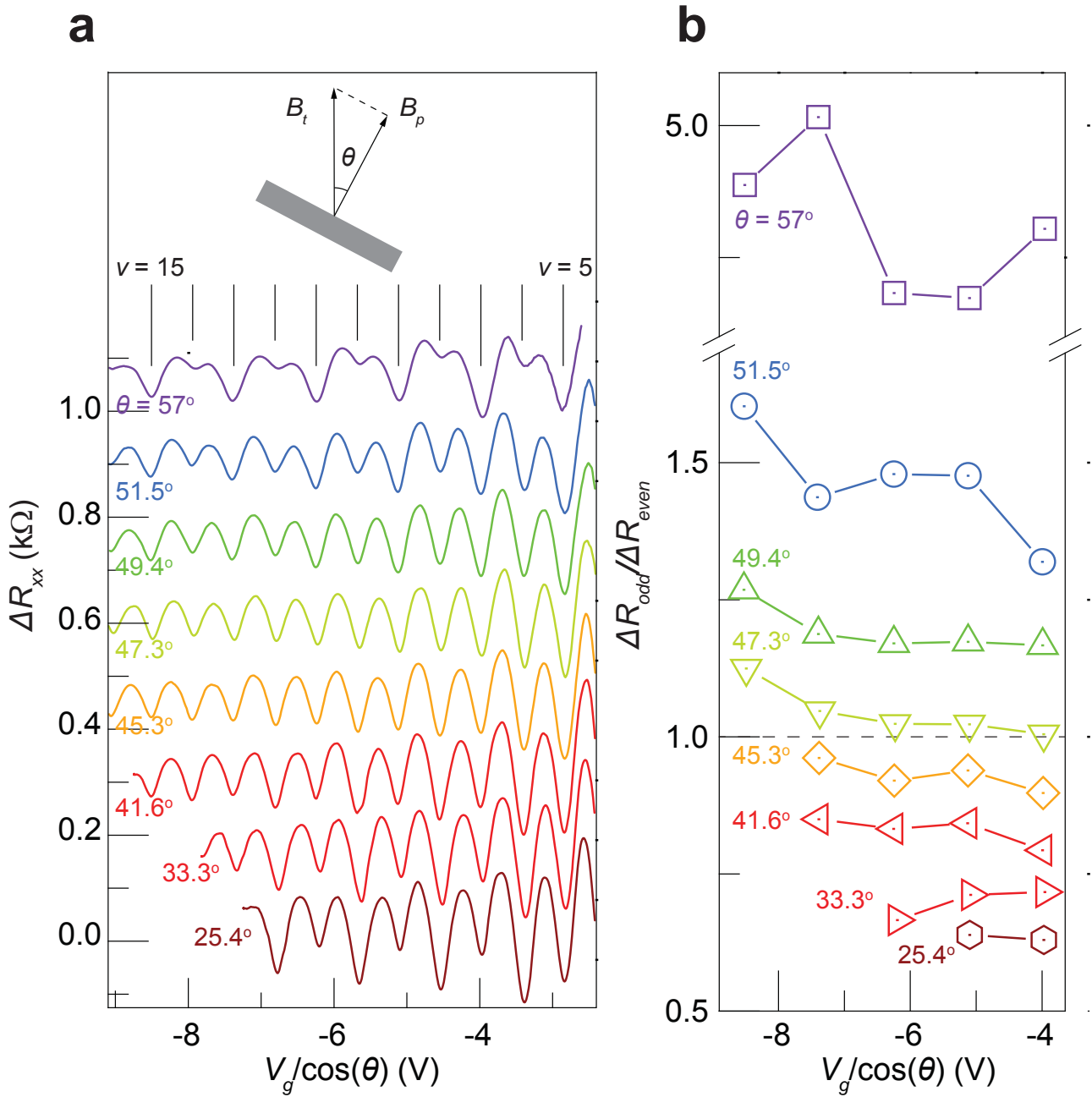

Likai Li *et al.*   Figure 4

Supplementary Information for

# Quantum Hall Effect in Black Phosphorus Two-dimensional Electron Gas

Likai Li, Fangyuan Yang, Guo Jun Ye, Zuocheng Zhang, Zengwei Zhu, Wen-Kai Lou, Liang Li, Kenji Watanabe, Takashi Taniguchi, Kai Chang, Yayu Wang, Xian Hui Chen[*] and Yuanbo Zhang[*]

*Email: zhyb@fudan.edu.cn, chenxh@ustc.edu.cn

**Content**

**1. Carrier density induced by the graphite back gate**

**2. Measurement of the quantum lifetime of holes**

**3. SdH oscillations at electron doping**

**4. Contact resistance of black phosphorus FET**

**5. Reported values of hole effective mass**

**6. Transport measurement in pulse magnetic field**

**7. Calculation of the *g*-factor**

**8. References**

# 1. Carrier density induced by the graphite back gate

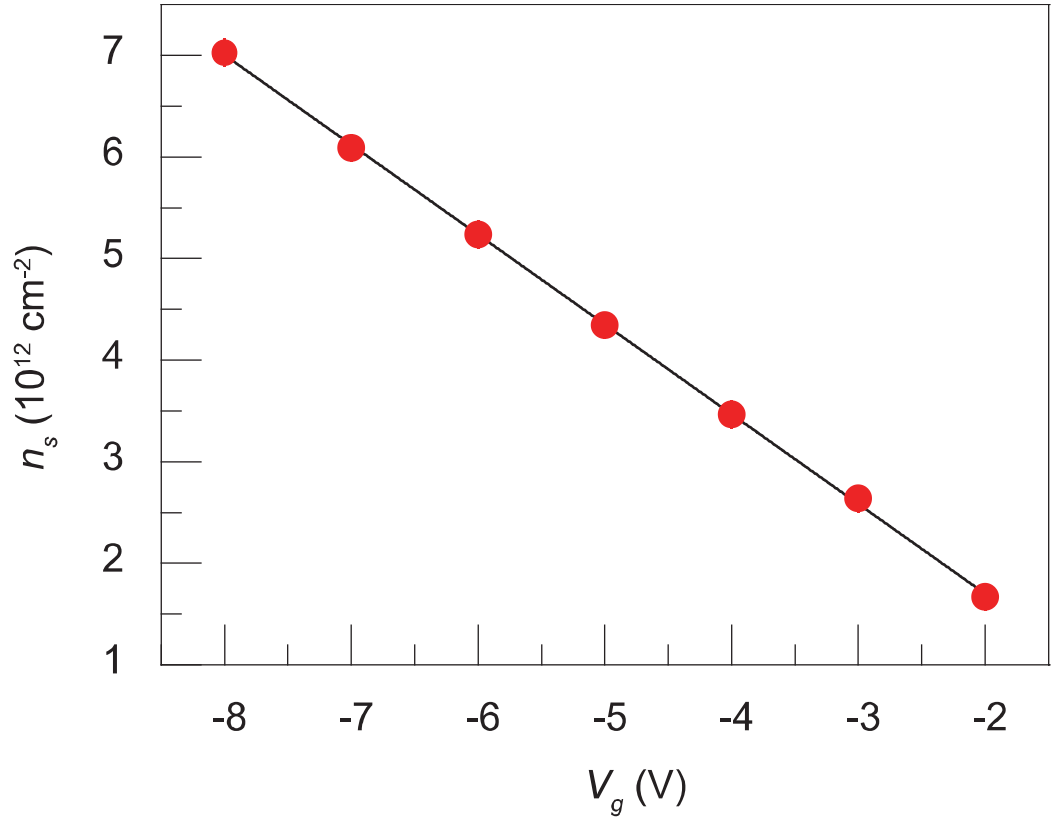

**Supplementary Figure 1 | Carrier density $n_H$ measured as a function of the gate voltage $V_g$.** $n_H$ values were obtained from Hall measurement. Line fit of the data (black line) yields a gate capacitance of $C_g = 8.84 \times 10^{11} e \text{ cm}^{-2} V^{-1}$. Such a gate capacitance implies an hBN thickness of $d \sim 25$ nm, if the parallel-capacitor model, $C_g = \varepsilon_0 \varepsilon_r / d$, is assumed. Here $\varepsilon_0$ is vacuum permittivity, and $\varepsilon_r \sim 4$ is the relative permittivity of hBN.

## 2. Measurement of the quantum lifetime of holes

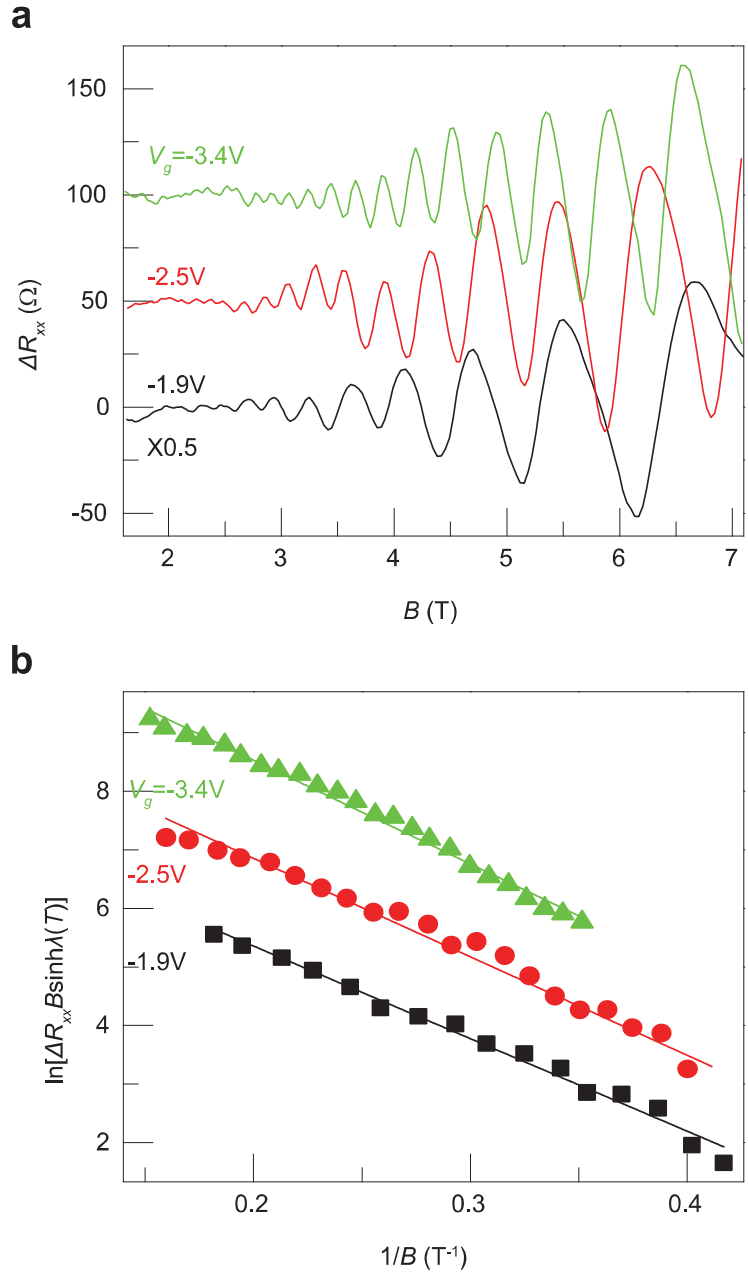

**Supplementary Figure 2 | Measurement of the quantum lifetime of holes. a,** Magnetoresistance (subtracted by a smooth background) as a function of magnetic field recorded at varying hole doping levels ($V_g < 0$). Curves are shifted vertically for clarity. SdH oscillations start to emerge at a critical magnetic field of $B_c \sim 2$ T. Data were obtained at $T = 0.3$ K. **b,** Dingle plots of $\ln[\Delta R_{xx} B \sinh\lambda(T)]$ versus $1/B$, where $\lambda(T) = 2\pi^2 k_B T m^*/\hbar eB$. The SdH oscillation amplitude $\Delta R_{xx}$ is extracted from **a**. Line fits of the Dingle plots (solid lines) yield an average quantum lifetime of 360 ps for the three hole doping levels under investigation.

## 3. SdH oscillations at electron doping

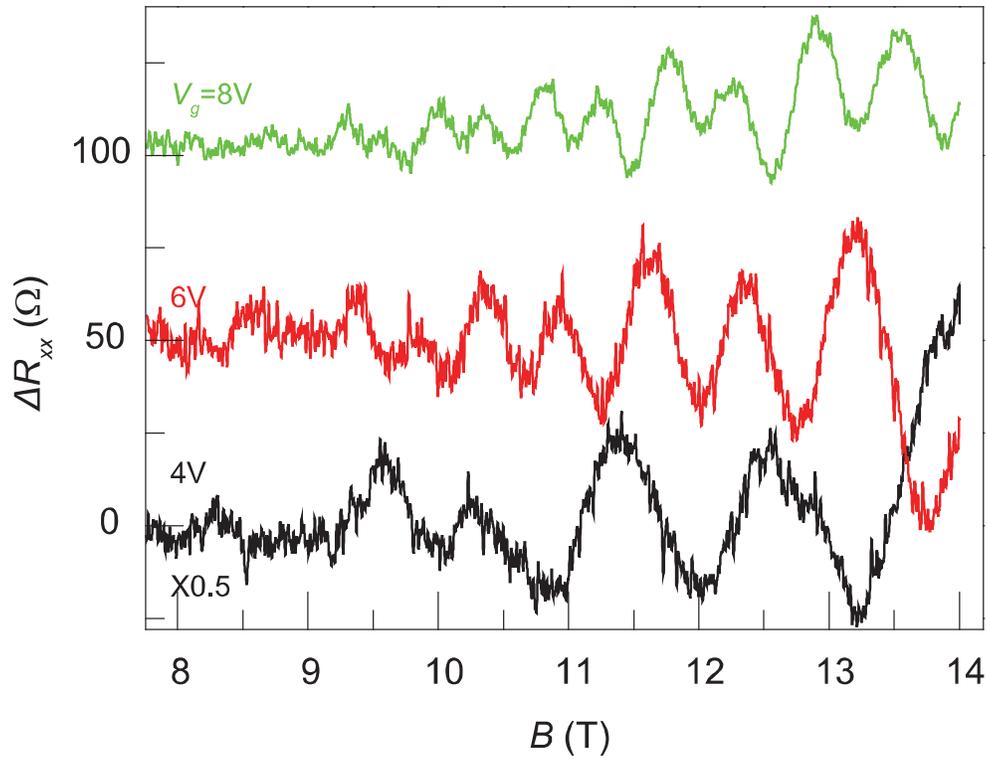

**Supplementary Figure 3 | SdH oscillations at electron doping.** Magnetoresistance (subtracted by a smooth background) is shown here as a function of magnetic field recorded at varying electron doping levels ($V_g > 0$). Curves are shifted vertically for clarity. Data were obtained at $T = 1.5$ K. The critical magnetic field, $B_c \sim 9$ T, marks the onset of SdH oscillations. Such a critical field is much lower than the value reported previously (15 T; ref. 26), and indicates a significant improvement on the electron mobility in our sample.

## 4. Contact resistance of black phosphorus FET

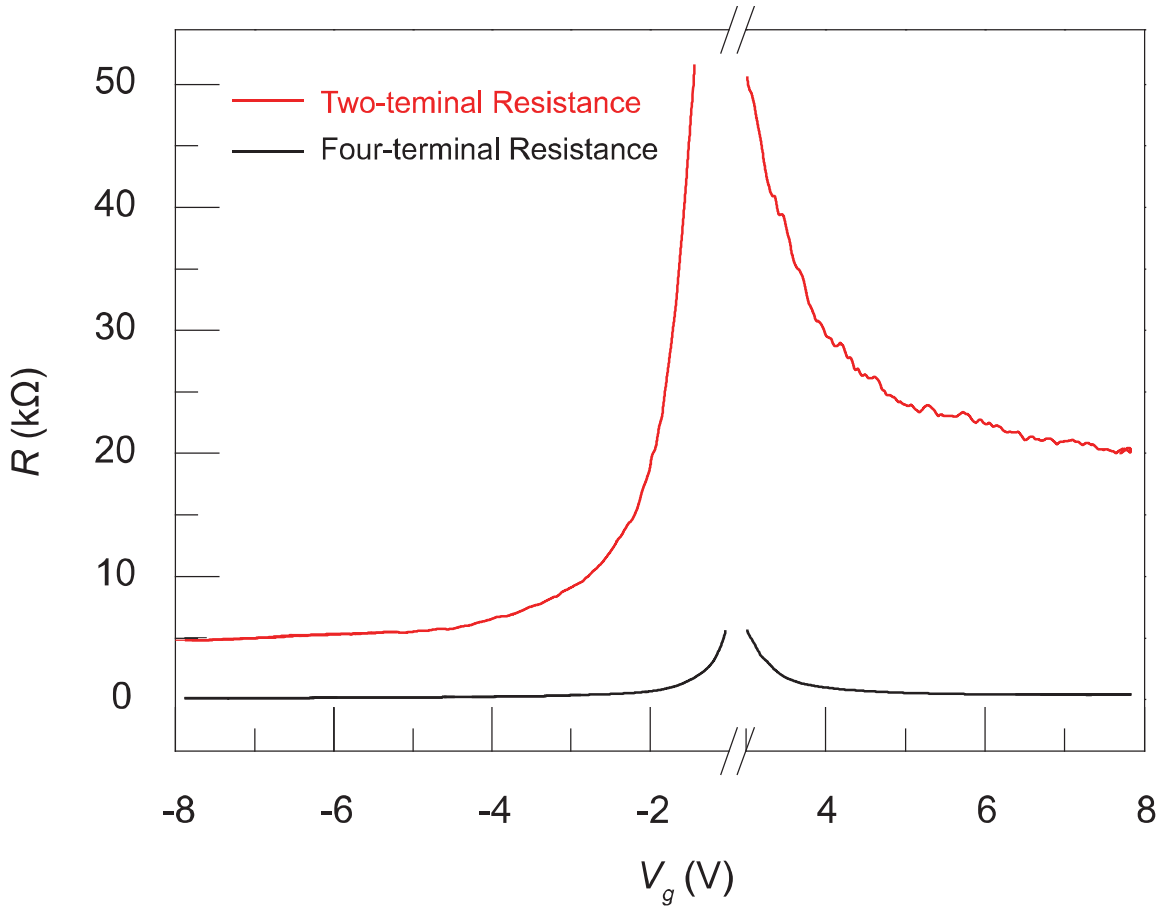

**Supplementary Figure 4 | Contact resistance of black phosphorus FET.** Two-terminal (red) and four-terminal resistance (black) is measured as a function of $V_g$ on the same sample discussed in Fig. 1, 2 and 3. The difference between the two resistances gives an order-of-magnitude estimate of the contact resistance. Data were obtained at $T = 1.5$ K. The large contact resistance at low hole doping levels ($> 10$ kΩ for $-2$ V $< V_g < 0$ V) makes it difficult to precisely measure the quantized $R_{xy}$ at $\nu = 1$ and 2 in high magnetic fields (Fig. 2b). Meanwhile, the contact resistance stays above 20 kΩ at electron doping ($V_g > 0$ V). Such a large contact resistance is probably due to the high Schottky barrier for electrons[1], and may have contributed to the fact that no quantum Hall effect was observed on the electron side.

## 5. Reported values of hole effective mass

| Hole Effective Mass $m^*/m_0$ | Magnetic Field $B$ (T) | Carrier Density $n_H$ ($10^{12}\ cm^{-2}$) | Source |
|---|---|---|---|
| 0.34±0.02 | 14.3 – 22.2 | 6.5 | ref. 26 |
| 0.26 – 0.31 | 12 | 5.1 – 2.5* | ref. 29 |
| 0.24±0.02 | 17 | 3.3 – 5.1 | ref. 30 |

**Supplementary Table 1 | Reported values of the effective mass of holes in black phosphorus 2DEG.**

* No carrier density value was provided in ref. 29. We estimate the carrier density from the frequency of the SdH oscillations shown in the paper.

## 6. Transport measurement in pulse magnetic field

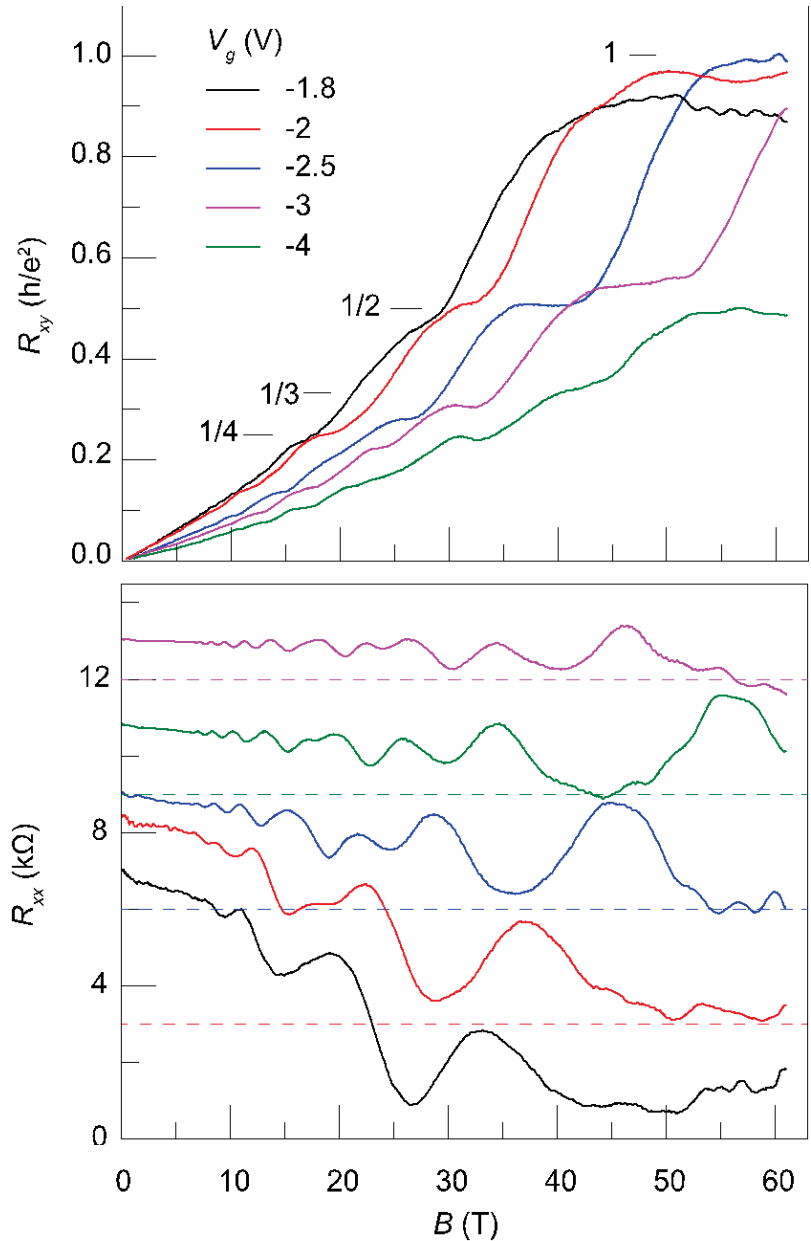

**Supplementary Figure 5 | Mageto-transport of holes in black phosphorus 2DEG in pulsed magnetic field.** Hall resistance (upper panel) and magnetoresistance (lower panel) are plotted as a function of magnetic field measured at varying hole doping levels ($V_g < 0$). Data were obtained in a pulsed magnetic field at $T = 1.5$ K. Horizontal lines and the fractions $1/\nu$ in the upper panel mark the quantized values in unit of $h/e^2$. Developing quantum Hall plateaus are observed at $\nu = 1$ and 2. Magnetoresistance curves are shifted vertically by multiples of 3 kΩ for clarity, and broken lines indicate zero-resistance level at each gate voltage.

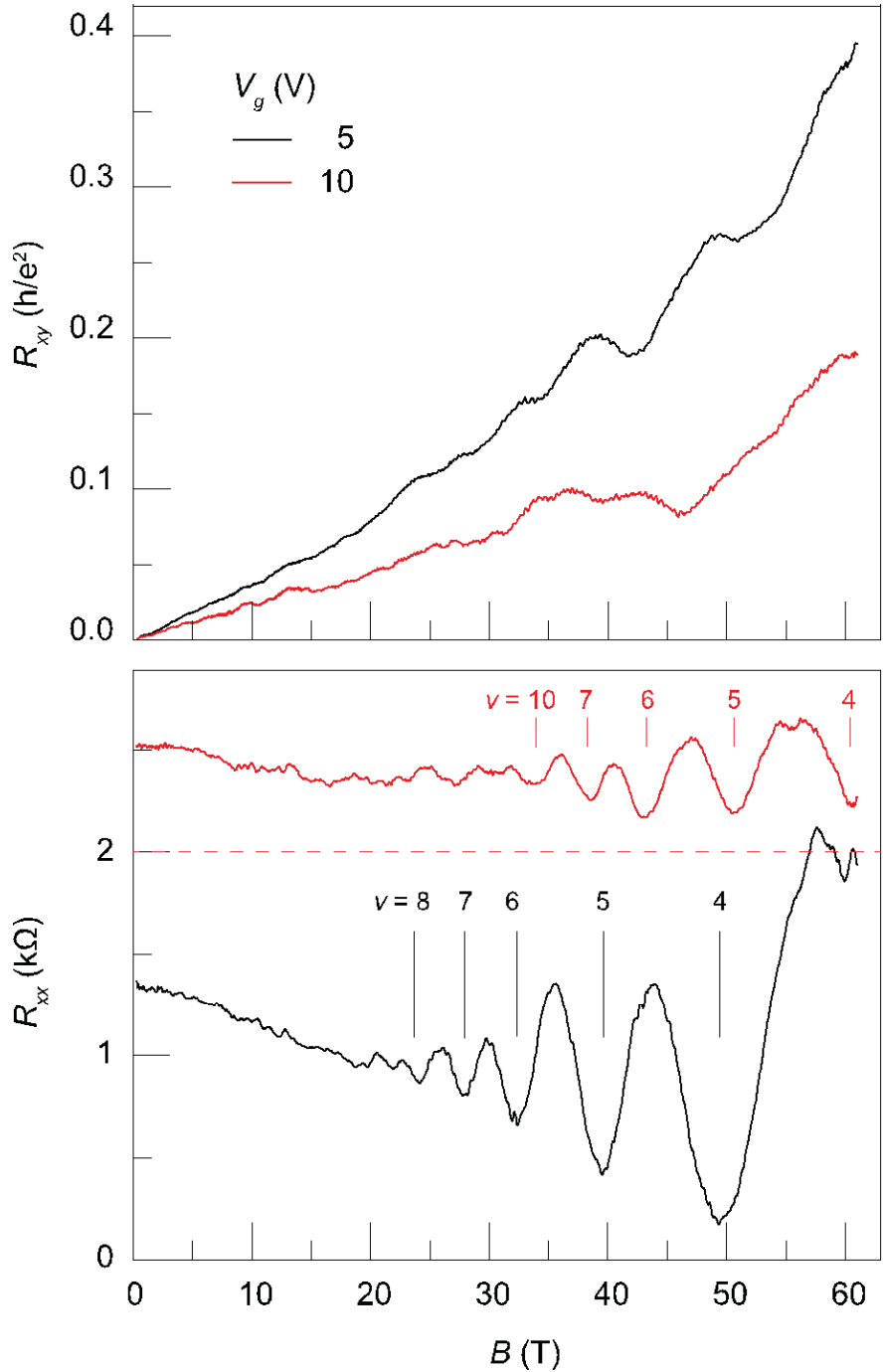

**Supplementary Figure 6 | Mageto-transport of electrons in black phosphorus 2DEG in pulsed magnetic field.** Hall resistance (upper panel) and magnetoresistance (lower panel) are plotted as a function of magnetic field measured at two different electron doping levels ($V_g > 0$). Data were obtained in a pulsed magnetic field at $T = 1.5$ K. Magnetoresistance at $V_g = -10$ V is shifted vertically by 2 kΩ for clarity. Vertical lines mark the filling factor $\nu$.

# 7. Calculation of the *g*-factor

### 7.1 Electronic structure of black phosphorus 2D quantum well

In this section we calculate the electronic structure of the gate-induced 2D quantum well at the surface of black phosphorus. The **k·p** Hamiltonian of the black phosphorus quantum well (QW) near $\Gamma$ point can be expressed as[S1]

$$H = \begin{bmatrix} E_c + \eta_c k_x^2 + v_c k_y^2 + \dfrac{\hbar^2}{2m_{cz}} \hat{k}_z^2 & \gamma k_x \\ \gamma k_x & E_v - \eta_v k_x^2 - v_v k_y^2 + \dfrac{\hbar^2}{2m_{cz}} \hat{k}_z^2 \end{bmatrix} + V(z),$$

where $V(z)$ describes the potential profile in the out-of-plane direction. $V(z)$ consists of the hard-wall confining potential at the sample surfaces and the internal electrostatic potential $V_{in}(z)$ caused by charge distribution in the QW. The subband dispersion and the corresponding eigenstates can be obtained numerically from the Schrödinger equation

$$H\psi_s = E_s \psi_s ,$$

where *s* is the subband index, and $\psi_s$ the envelope function.

To solve the Schrödinger equation, we adopt the hard-wall boundary condition and expand $\psi_s$ as $\psi_s = \sqrt{\dfrac{2}{L}} \sum_m \sin\left(\dfrac{m\pi}{L}\right)$, where *L* is the width of the quantum well. The internal electrostatic potential $V_{in}(z)$ is determined by the Poisson equation

$$\frac{d^2 V_{in}(z)}{dz^2} = -\frac{[n(z) + p(z)]}{\varepsilon} ,$$

where $n(z)$ and $p(z)$ are the densities of electrons and holes in *z* direction, respectively, and $\varepsilon$ is the dielectric constant. $n(z)$ and $p(z)$ can be obtained from[S2]

$$n(z) = \sum_i n_i |\psi_i^c(z)|^2$$
$$p(z) = \sum_i p_i |\psi_i^v(z)|^2 ,$$

where *c* and *v* refer to the conduction and valence bands, respectively, and

$$\begin{cases} n_i = -m_i^{c*}|e_0|\dfrac{k_B T}{\pi \hbar^2}\ln\left(e^{\frac{E_f - E_i^c}{k_B T}} + 1\right) \\ p_i = m_i^{v*}|e_0|\dfrac{k_B T}{\pi \hbar^2}\ln\left(e^{\frac{E_i^v - E_f}{k_B T}} + 1\right) \end{cases}.$$

Here $E_f$ is the Fermi energy, and $m^*$ refers to the effective mass given by $\sqrt{m_x^* m_y^*}$ (ref S3). We obtain the eigenstates and eigenenergies of the QW numerically by solving the Schrödinger and Poisson equation self-consistently. Because the samples under investigation are p-type, we only need to consider the electric static potential induced by the hole states in numerical calculations.

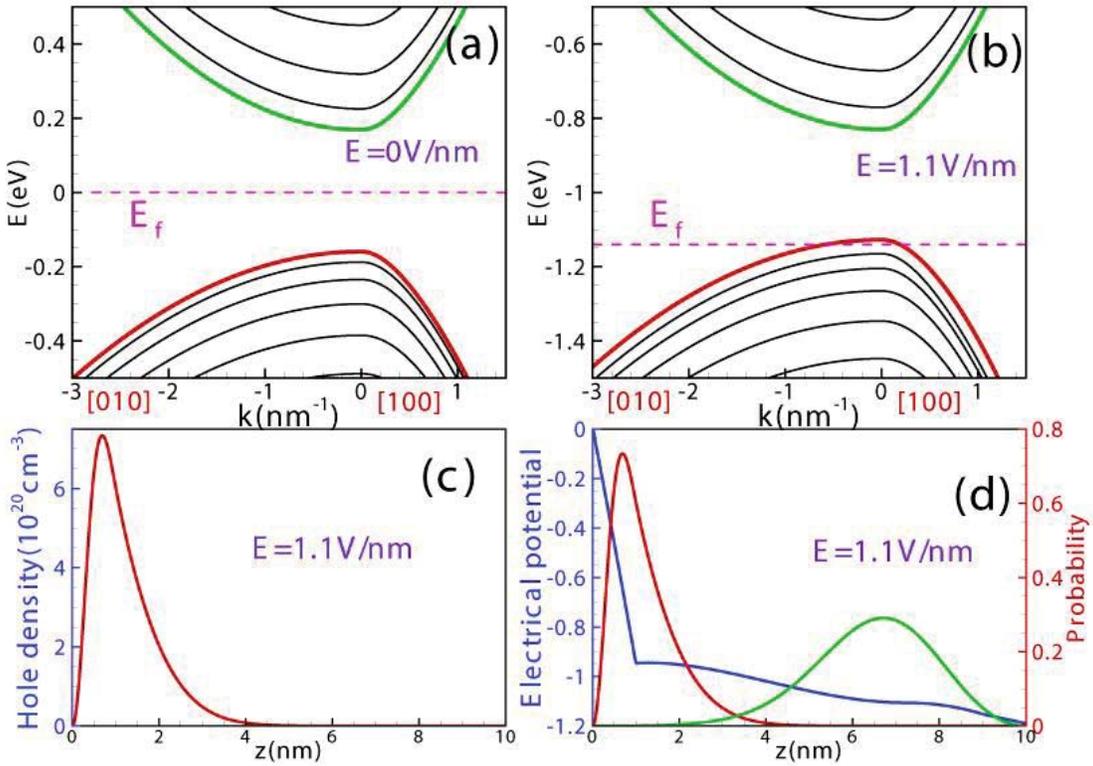

**Supplementary Figure 7 | Electronic structure of gate-induced black phosphorus QW. a,** Band structure of a 10-nm-thick pristine black phosphorus without gate electric field. The pink dashed line indicates the Fermi level. **b,** Band structure of black phosphorus QW induced by a gate electric field of $E = 1.1$ V/nm. System geometry is the same as in **a**. **c,** Spatial distribution of hole density at the surface of black phosphorus under a gate electric field of $E = 1.1$ V/nm. **d,** Spatial distributions of the wavefunction of the highest hole subband (red) and the lowest electron subband (green). Blue curve denotes the electric potential induced by the gate electric field.

The band structure of black phosphorus QW without and with gate electric field is shown in Figs. 1a and 1b, respectively. Here the thickness of the black phosphorus sample is assumed to be 10 nm and the gate electric field at the sample surface is $E = 1.1$ V/nm. Both values are close to those used in the experiment. Other parameters used in the calculation are given in Table S2. Our results indicate that only the highest hole subband is occupied and the energy gap in the QW remains ~ 0.3 eV. Our calculation further shows that free holes are mostly confined within ~ 2 atomic layers at the surface, as shown in Fig. S7c.

| $E_c$ (eV) | $E_v$ (eV) | $\eta_c$ (eV nm$^2$) | $\nu_c$ (eV nm$^2$) | $\gamma$ (eV nm) |
|---|---|---|---|---|
| 0.155 | -0.155 | 0.191 | 0.0545 | 0.284 |
| $m_{cz}$ ($m_0$) | $m_{cz}$ ($m_0$) | $\eta_v$ (eV nm$^2$) | $\nu_v$ (eV nm$^2$) | $\varepsilon$ ($\varepsilon_0$) |
| 0.2 | 0.4 | 0.191 | 0.0381 | 10.0 |

**Supplementary Table 2 | Parameters used in the calculation.** $\varepsilon_0$ is vacuum permittivity.

### 7.2 *g*-factor of the 2DEG in black phosphorus quantum well

Armed with the electronic structure of the QW, we now calculate the *g*-factor of the 2DEG based on a multiband **k·p** theory. Within the **k·p** framework, the Schrödinger equation can be expressed as[S4]

$$\sum_{n'} \left( \sum_{\alpha\beta} D_{nn'\alpha\beta} k_\alpha k_\beta - E\delta_{nn'} \right) \psi_{n'} = 0,$$

Here the Hamiltonian matrix element coefficients are given by

$$D_{nn'\alpha\beta} = \frac{\hbar^2}{2m_0} \delta_{nn'} \delta_{\alpha\beta} + \frac{\hbar^2}{m_0^2} \sum_l \frac{\langle n|P_\alpha|l\rangle\langle l|P_\beta|n'\rangle}{E_n^{(0)} - E_l^{(0)}} \quad \text{with} \quad \vec{P} = \vec{p} + \frac{\hbar}{4m_0 c^2} \vec{\sigma} \times \nabla V,$$

so the secular equation can be written as

$$\sum_{n'} \left( \sum_{\alpha\beta} \frac{1}{2} \left[ D^S_{nn'\alpha\beta} \{k_\alpha, k_\beta\} + D^A_{nn'\alpha\beta} [k_\alpha, k_\beta] \right] - E\delta_{nn'} \right) \psi_{n'} = 0 \quad (1)$$

where $D^S_{nn'\alpha\beta}$ and $D^A_{nn'\alpha\beta}$ are the symmetric and antisymmetric part of $D_{nn'\alpha\beta}$ (i.e., $D^S_{nn'\alpha\beta} = (D_{nn'\alpha\beta} + D_{nn'\beta\alpha})/2$ and $D^A_{nn'\alpha\beta} = (D_{nn'\alpha\beta} - D_{nn'\beta\alpha})/2$), respectively. $[k_\alpha, k_\beta]$ and $\{k_\alpha, k_\beta\}$ are commutator and anticommutator, respectively. $D^S_{nn'\alpha\beta}$ can be written as

$$D^{S}_{nn'\alpha\beta} = \frac{\hbar^2}{2m_0}\delta_{nn'}\delta_{\alpha\beta} + \frac{\hbar^2}{2m_0^2}\sum_l \frac{\langle n|P_\alpha|l\rangle\langle l|P_\beta|n'\rangle + \langle n|P_\beta|l\rangle\langle l|P_\alpha|n'\rangle}{E_n^{(0)} - E_l^{(0)}},$$

and the antisymmetric part $D^{S}_{nn'\alpha\beta}$ is given by

$$D^{A}_{nn'\alpha\beta} = \frac{\hbar^2}{2m_0^2}\sum_l \frac{\langle n|P_\alpha|l\rangle\langle l|P_\beta|n'\rangle - \langle n|P_\beta|l\rangle\langle l|P_\alpha|n'\rangle}{E_n^{(0)} - E_l^{(0)}}.$$

When the system is subjected to a magnetic field, the secular equation becomes

$$\sum_{n'}\left(\sum_{\alpha\beta}\frac{1}{2}\left[D^{S}_{nn'\alpha\beta}\{k_\alpha,k_\beta\} + D^{A}_{nn'\alpha\beta}[k_\alpha,k_\beta]\right] + \mu_B\vec{\sigma}\cdot\vec{B} - E\delta_{nn'}\right)\psi_{n'} = 0 \quad (2)$$

where Bohr magneton $\mu_B = |e|\hbar/(2m_0)$, $[k_\alpha,k_\beta] = \varepsilon_{\alpha\beta\gamma}eB_\gamma/(i\hbar)$, and $\varepsilon_{\alpha\beta\gamma}$ is the Levi-Civita symbol. We consider an external magnetic field applied along $z$ axis. $\hbar\vec{k}$ is now replaced by the canonical momentum, $\hbar\vec{k} \to \hbar\vec{k} + e\vec{A}$, where $\vec{A} = B_z(-y,x,0)/2$ is the vector potential adopting the symmetry gauge. In this case,

$$[k_x,k_y] = \mu_B\frac{2m_0}{i\hbar^2}B_z.$$

So the secular equation (2) becomes

$$\sum_{n'}\left(\begin{array}{l}\sum_{\alpha\beta}\frac{1}{2}D^{S}_{nn'\alpha\beta}\{k_\alpha,k_\beta\} \\ +\mu_B\frac{1}{2im_0}\sum_l \frac{\langle n|P_\alpha|l\rangle\langle l|P_\beta|n'\rangle - \langle n|P_\beta|l\rangle\langle l|P_\alpha|n'\rangle}{E_n^{(0)} - E_l^{(0)}}B_z + \mu_B m_s B_z \\ -E\delta_{nn'}\end{array}\right)\psi_{n'} = 0 \quad (3)$$

From equation (3) we obtain the analytical form of the effective magnetic moment of the electron in a crystal:

$$\mu^* = \mu_B\left[\delta_{\alpha\beta} + \frac{1}{im_0}\sum_l \frac{\langle n|P_\alpha|l\rangle\langle l|P_\beta|n\rangle - \langle n|P_\beta|l\rangle\langle l|P_\alpha|n\rangle}{E_n^{(0)} - E_l^{(0)}}\right],$$

so the *g*-factor is give by

$$g^* = 2\mu^*/\mu_B = g_0\left[\delta_{\alpha\beta} + \frac{1}{im_0}\sum_l \frac{\langle n|P_\alpha|l\rangle\langle l|P_\beta|n\rangle - \langle n|P_\beta|l\rangle\langle l|P_\alpha|n\rangle}{E_n^{(0)} - E_l^{(0)}}\right].$$

where $g_0 = 2$ is the electron $g$-factor.

It has been shown that the dominant components of electrons states near $\Gamma$ point are $p_z$ orbitals[S5,S6]. $\langle l|P_z|\Gamma_{2v}^+, \uparrow\rangle$ is therefore very small, and can be safely neglected. So

$$g_{xx} = g_0 \left[1 + \frac{1}{im_0} \sum_l \frac{\langle \Gamma_{2v}^+, \uparrow|P_y|l\rangle\langle l|P_z|\Gamma_{2v}^+, \uparrow\rangle - \langle \Gamma_{2v}^+, \uparrow|P_z|l\rangle\langle l|P_y|\Gamma_{2v}^+, \uparrow\rangle}{E_{\Gamma_{2v}^+,\uparrow}^{(0)} - E_l^{(0)}}\right] \simeq g_0,$$

$$g_{yy} = g_0 \left[1 + \frac{1}{im_0} \sum_l \frac{\langle \Gamma_{2v}^+, \uparrow|P_z|l\rangle\langle l|P_x|\Gamma_{2v}^+, \uparrow\rangle - \langle \Gamma_{2v}^+, \uparrow|P_x|l\rangle\langle l|P_z|\Gamma_{2v}^+, \uparrow\rangle}{E_{\Gamma_{2v}^+,\uparrow}^{(0)} - E_l^{(0)}}\right] \simeq g_0,$$

and

$$g_{zz} = g_0 \left[1 + \frac{1}{im_0} \sum_l \frac{\langle \Gamma_{2v}^+, \uparrow|P_x|l\rangle\langle l|P_y|\Gamma_{2v}^+, \uparrow\rangle - \langle \Gamma_{2v}^+, \uparrow|P_y|l\rangle\langle l|P_x|\Gamma_{2v}^+, \uparrow\rangle}{E_{\Gamma_{2v}^+,\uparrow}^{(0)} - E_l^{(0)}}\right]$$

$$= g_0 \left[1 + \frac{1}{im_0} \begin{bmatrix} \sum_{\Gamma_{4c}^-,\uparrow} \frac{\langle \Gamma_{2v}^+, \uparrow|P_x|\Gamma_{4c}^-, \uparrow\rangle\langle \Gamma_{4c}^-, \uparrow|P_y|\Gamma_{2v}^+, \uparrow\rangle}{E_{\Gamma_{2v}^+,\uparrow}^{(0)} - E_{\Gamma_{4c}^-,\uparrow}^{(0)}} \\ - \sum_{\Gamma_{4c}^-,\uparrow} \frac{\langle \Gamma_{2v}^+, \uparrow|P_y|\Gamma_{4c}^-, \uparrow\rangle\langle \Gamma_{4c}^-, \uparrow|P_x|\Gamma_{2v}^+, \uparrow\rangle}{E_{\Gamma_{2v}^+,\uparrow}^{(0)} - E_{\Gamma_{4c}^-,\uparrow}^{(0)}} \\ + \sum_{\Gamma_{1v}^-,\uparrow} \frac{\langle \Gamma_{2v}^+, \uparrow|P_x|\Gamma_{1v}^-, \uparrow\rangle\langle \Gamma_{1v}^-, \uparrow|P_y|\Gamma_{2v}^+, \uparrow\rangle}{E_{\Gamma_{2v}^+,\uparrow}^{(0)} - E_{\Gamma_{1v}^-,\uparrow}^{(0)}} \\ - \sum_{\Gamma_{1v}^-,\uparrow} \frac{\langle \Gamma_{2v}^+, \uparrow|P_y|\Gamma_{1v}^-, \uparrow\rangle\langle \Gamma_{1v}^-, \uparrow|P_x|\Gamma_{2v}^+, \uparrow\rangle}{E_{\Gamma_{2v}^+,\uparrow}^{(0)} - E_{\Gamma_{1v}^-,\uparrow}^{(0)}} \end{bmatrix}\right].$$

In the QW, the total wave function $\Phi$ is the product of envelope function $\psi$ and band-edge Bloch function $\Gamma$, so $|\Phi_i\rangle = |\Gamma_i\rangle \otimes |\psi_i\rangle$. The above expression for $g_{zz}$ can then be written as

$$g_{zz} = g_0 \left[ 1 + \frac{1}{im_0} \left[ \begin{array}{c} \sum_{\Gamma_{4c}^-,\uparrow} \dfrac{\langle \Phi_{2v}^+,\uparrow |P_x| \Phi_{4c}^-,\uparrow \rangle \langle \Phi_{4c}^-,\uparrow |P_y| \Phi_{2v}^+,\uparrow \rangle}{E^{(0)}_{\Gamma_{2v}^+,\uparrow} - E^{(0)}_{\Gamma_{4c}^-,\uparrow}} \\ -\sum_{\Gamma_{4c}^-,\uparrow} \dfrac{\langle \Phi_{2v}^+,\uparrow |P_y| \Phi_{4c}^-,\uparrow \rangle \langle \Phi_{4c}^-,\uparrow |P_x| \Phi_{2v}^+,\uparrow \rangle}{E^{(0)}_{\Gamma_{2v}^+,\uparrow} - E^{(0)}_{\Gamma_{4c}^-,\uparrow}} \\ +\sum_{\Gamma_{1v}^-,\uparrow} \dfrac{\langle \Phi_{2v}^+,\uparrow |P_x| \Phi_{1v}^-,\uparrow \rangle \langle \Phi_{1v}^-,\uparrow |P_y| \Phi_{2v}^+,\uparrow \rangle}{E^{(0)}_{\Gamma_{2v}^+,\uparrow} - E^{(0)}_{\Gamma_{1v}^-,\uparrow}} \\ -\sum_{\Gamma_{1v}^-,\uparrow} \dfrac{\langle \Phi_{2v}^+,\uparrow |P_y| \Phi_{1v}^-,\uparrow \rangle \langle \Phi_{1v}^-,\uparrow |P_x| \Phi_{2v}^+,\uparrow \rangle}{E^{(0)}_{\Gamma_{2v}^+,\uparrow} - E^{(0)}_{\Gamma_{1v}^-,\uparrow}} \end{array} \right] \right]$$

$$\simeq g_0 \left[ 1 + \frac{2\alpha_{42} P_{x_1} \langle \psi_{2v}^+ | \psi_{4c}^- \rangle}{\hbar E_g} \right]$$

where

$$P_{x_1} = \langle \Gamma_{4c}^-,\uparrow |P_x| \Gamma_{2v}^+,\uparrow \rangle,$$

$$\alpha_{42} = \sum_{\Gamma_{4c}^-,\uparrow} \frac{i\hbar^2}{4m_0^2 c^2} \langle \Gamma_{4c}^-,\uparrow \left| \frac{\partial V}{\partial x} \right| \Gamma_{2v}^+,\uparrow \rangle.$$

From our calculation shown in Fig. S7d, one clearly sees that the gate electrical field pushes electrons and holes in opposite directions, so the envelope wavefunctions of the electron and hole states (red and green in Fig. S7d, respectively) are spatially separated. The overlap integral $\langle \psi_{2v}^+ | \psi_{4c}^- \rangle$ therefore becomes negligibly small, so $g_{zz} \simeq g_0$.

In summary, the $g$-factor of the hole carriers in the gate-induced 2DEG in black phosphorus is $g \approx 2$. In addition, the anisotropy of the $g$-factor is negligible (*i.e.*, $g_{xx} \simeq g_{yy} \simeq g_{zz}$) as a result of the weak intrinsic spin-orbit interaction in black phosphorus.